\documentclass[non-peer-review]{fa2026}

\usepackage[T1]{fontenc}
\usepackage{ifthen}
\usepackage{booktabs}
\usepackage{array}

\title{Frequency-Modulated and Single-Tone Excitation to Reveal Vibro-Acoustic Nonlinearities in Loosened Bolted Joints}

\author[1]{Berkay Kullukcu}
\author[1]{Robin Pianowski}
\author[1]{Dina Hannebauer}
\correspondingauthor{berkay.kullukcu@th-wildau.de}{Berkay Kullukcu et al.}

\affil[1]{FG Maschinendynamik und Akustik, Technische Hochschule Wildau, 15745 Brandenburg, Germany}

\begin{document}
\maketitle

\begin{abstract}
Preload loss in bolted joints results in alterations of the stiffness, damping, and nonlinearity of the structure, but existing monitoring techniques for rail-vehicle systems are often not capable of combining controlled shaker tests and sensing of nonlinear features. This paper proposes a method for detecting bolt loosening using a vibro-acoustic technique, where the structure is subjected to controlled shaker tests to sense the nonlinear features. A triaxial accelerometer was attached to the demonstrator, a microphone was placed in close proximity, and one of the bolts was tested under 0\%, 20\%, 40\%, and 80\% preload conditions. Single-tone and frequency-modulated (FM) signals close to the main natural frequency of 130 Hz, which was identified using sine sweep and narrow-band excitation, were applied to the demonstrator. When the structure was subjected to 130 Hz single-tone excitation, the loose state of the bolt exhibited several additional high-frequency spectral peaks. FM excitation between 125 and 135 Hz further distinguished between the states. Harmonic band power ratios, normalized to the carrier, distinguished between the loose state and the 80\% preload state, where the difference between the loose and 80\% preload states was 17.5 dB for $l = 2$ and 36.5 dB for $l = 6$.
\end{abstract}


\section{Introduction}\label{sec:introduction}

Bolted joints are commonly used due to their high strength, demountability, and low cost, but they need to be subjected to a preload or clamp force to keep the parts in frictional contact with each other. If the preload is reduced, the joints become loose and start to rattle, losing their stiffness and sometimes accumulating high levels of fatigue. For rail vehicles, this means potential maintenance issues and, in extreme cases, safety-related degradation of the interface subjected to repeated cycles of vibration and shock. In other words, ``a tight bolt'' means a stiff spring with frictional damping properties, while ``a loose bolt'' means a soft and nonlinear contact with unpredictable damping properties. As literature proves: ``Even with small reductions in preload in a bolted plate assembly, resonances will be affected and noise radiation will be increased before the nut is rotated noticeably.'' [1], [2], [3]

Rail vehicles have complex patterns of bolted interfaces on the carbody and roofborne structures, which are typically exposed to broadband excitation from track irregularities, curving, and braking/traction events with a tendency to increase transverse motion at the interface. This causes increased rattle and structure-borne noise, altered transference of vibrations, and an increased rate of damage at the interface, although a lack of access makes it difficult to retrofit preload sensors on in-service vehicles. The application of a shaker, a small number of accelerometers, and a microphone is well adapted to this problem, as it is based on measurable changes in dynamic response and sound radiation capabilities that can be achieved with access to in-service vehicles and can be adapted to examine the key resonant bands in body vibration problems and rail vehicle maintenance. [4], [5]--[8]

If transverse excitation on the interface is sufficient to produce micro-slip and/or macro-slip, then thread friction can be overcome and rotation of the bolt can occur, leading to loss of preload. This can typically be shown to be the case in the ``Junker'' transverse vibration configuration. If a lap joint is exposed to transverse excitation, then the two plates will stick and then slip past each other, and each slip event will cause a small amount of unwinding. [2], [4], [9], [10]

The test setup for transverse vibrations for both DIN 65151 and ISO 16130 have already been standardized. These standards also define the procedure for comparison and quantification of the clamp force decay under controlled displacement excitation. However, in most monitoring situations, there is no access to the preload/axial force. For example, when working with an existing vehicle or demonstrator, it is easy to attach an accelerometer and position a microphone. However, it is not possible to replace every bolt with a sensing bolt. [9]--[11]

The preload state changes the interface state by altering (i) the effective stiffness, (ii) the amount of frictional energy loss (damping), and (iii) the degree of nonlinearity. Research into jointed structures indicates that the primary source of damping at bolted interfaces is caused by micro-slip. For example, when the preload decreases, the interface state changes from sticking to partially slipping, which becomes nonlinear. This results in harmonics and sidebands in the frequency spectrum. [12]--[18]

Many studies have already successfully employed various instrumentation schemes, such as strain-sensing bolts, ultrasonic time-of-flight, EMI with piezoelectric patches, and wave propagation. These have proven to be highly sensitive to changes. However, they also add to the complexity. On the other hand, accelerometers and microphones are easy to use and inexpensive. However, their use for quantifying loosening events is less developed. For example, an accelerometer can be fixed to a bracket for easy mounting, and a microphone can be placed at a fixed distance from the interface. This allows repeated experiments with minimal changes to the setup. [2], [19]

One of the most common concepts is to analyze the changes in modal parameters or frequency response functions. Modal testing theory offers a direct relationship between frequency response functions and physical parameters. Recent research has demonstrated the ability of even limited accelerometer measurements and machine learning to detect bolt tightness changes. For example, for a swept-sine shaker input, the loosened joint might decrease the peak frequency of the resonance peak and increase the bandwidth of the peak. This can be achieved with peak tracking or frequency response function distance methods. [20], [21]--[26]

Percussive and vibration-driven sound changes have also been researched for the detection of bolt looseness. The sound field generated in this case can be considered a measurable ``shadow'' of the vibration. Various studies have employed approaches ranging from conventional feature engineering to classification and feature selection for effective detection. Some studies have also attempted to relate the sound changes to the actual vibration. For example, the ring-down sound generated in a truss connection due to a strike or other means is a characteristic sound field. With increasing looseness, the ring-down time increases or decreases for various frequencies. [27]--[34]

For a system like the bolt interface, where there is a transition from sticking to slipping, there exists a nonlinear modulator. Contact acoustic nonlinearity (CAN) and vibro-acoustic modulation (VAM) have attempted to utilize this phenomenon by detecting higher-order harmonics and sidebands that increase in amplitude with increasing looseness. While the initial approach utilizes a high-frequency signal for this detection, the same phenomenon can be utilized for a shaker-based approach and various frequencies in the audio range for the detection of harmonic distortion and modulation products. For example, a low-frequency tone applied by a shaker to a loose connection will have additional frequency components at integer multiples of the tone (harmonics) or integer multiples of a second frequency (sidebands) in the acceleration and sound spectra. [35]--[38]

Current standards and many loosening-focused studies focus on measuring the rate of clamp force reduction with a specified transverse motion, while many studies on structural health monitoring deal with classification under a specified excitation or employ particular transducers. There is less research on systems that combine a realistic bolted-joint problem, controlled shaker inputs over a range of excitation frequencies and amplitudes, simultaneous measurements of vibrations and sound, and a set of interpretable linear and nonlinear measures that can be derived from a combination of accelerometers and a microphone. The aim is to determine the excitation regimes where loosening is probable for the problem and to develop a minimal-sensor recognition process that remains informative in the absence of direct preload measurements. [9]--[11], [23], [27], [35]

\section{Method}\label{sec:method}

The work employed a demonstrator as given in Fig.~1. The vibration signals were introduced to the demonstrator using a mechanical shaker (TMS K2007E01), connected by a plastic rod to a force and acceleration sensor (PCB 288D01) using a threaded connection. Another three-axis accelerometer (PCB 356A01/NC) was also placed on top of the demonstrator. A microphone was placed at a distance of 300 mm from the bolt under test.

\begin{figure}[t]
  \centering
  \includegraphics[width=0.65\columnwidth]{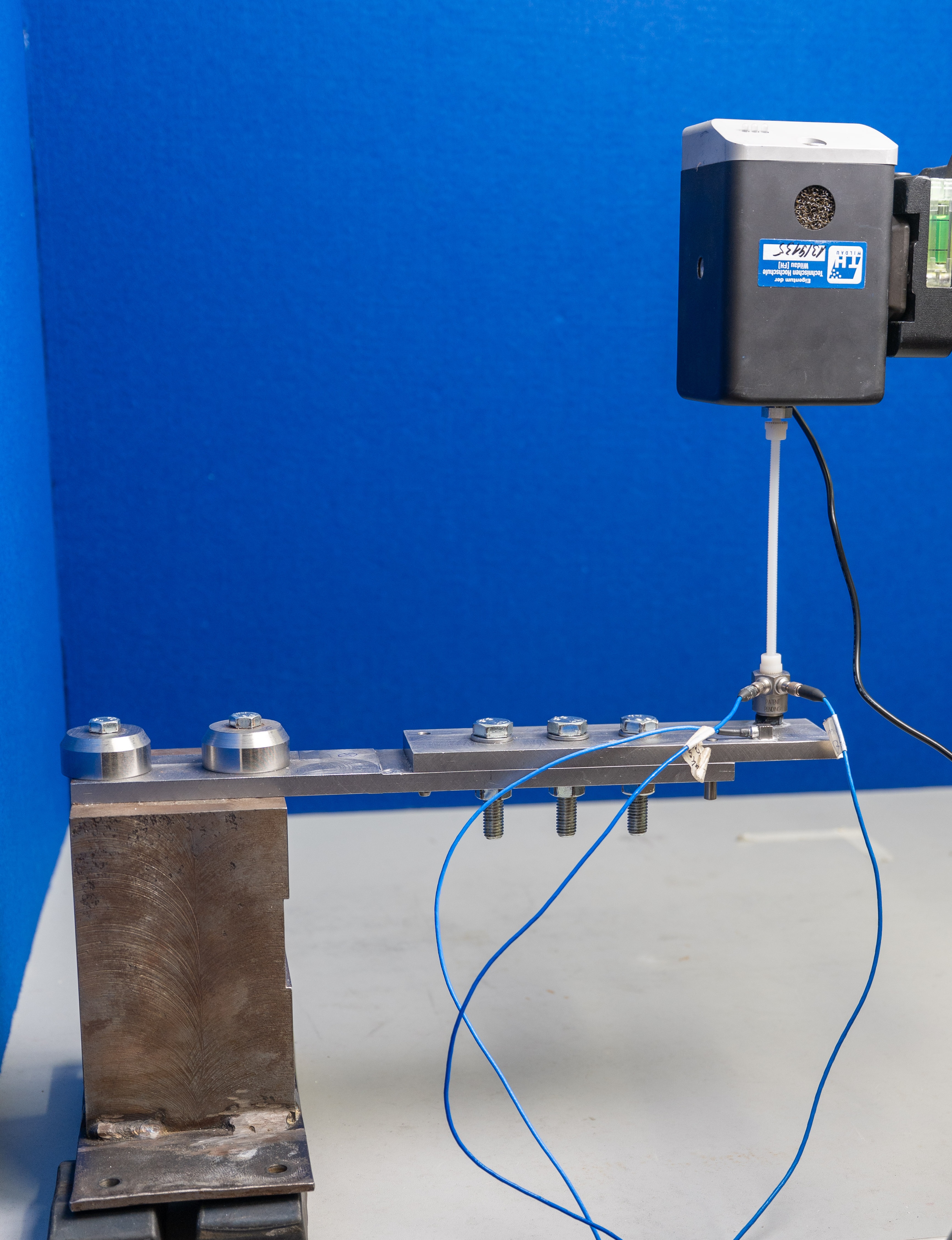}
  \caption{Experimental setup: rail-inspired bolted demonstrator excited by electrodynamic shaker (TMS K2007E01) via force sensor (PCB 288D01). Triaxial accelerometer (PCB 356A01) monitors structural response; a microphone positioned 300 mm from bolt S2 (under test) captures radiated sound. The configuration enables non-invasive vibro-acoustic monitoring without preload sensors.}
  \label{fig:setup}
\end{figure}

The bolt under test was tightened to have 0\%, 20\%, 40\%, and 80\% preload, corresponding to 0 Nm, 12.5 Nm, 25 Nm, and 50 Nm, respectively. The two M8 bolts connecting the cantilever beams to the metal block were tightened to 80\% preload (25 Nm). Three M10 bolts connected the two metal beams together with a nut and washers on both sides. Only bolt S2 (Fig.~2) had its tightening torque changed, whereas bolts S1 and S3 remained at 80\% preload throughout the experiment. For each preload state, the mechanical shaker applied single-tone and frequency-modulated vibrations to the demonstrator. FM vibrations were used to understand the effect of frequency modulation on the modal behavior of the system.

\begin{figure}[t]
  \centering
  \includegraphics[width=0.98\columnwidth]{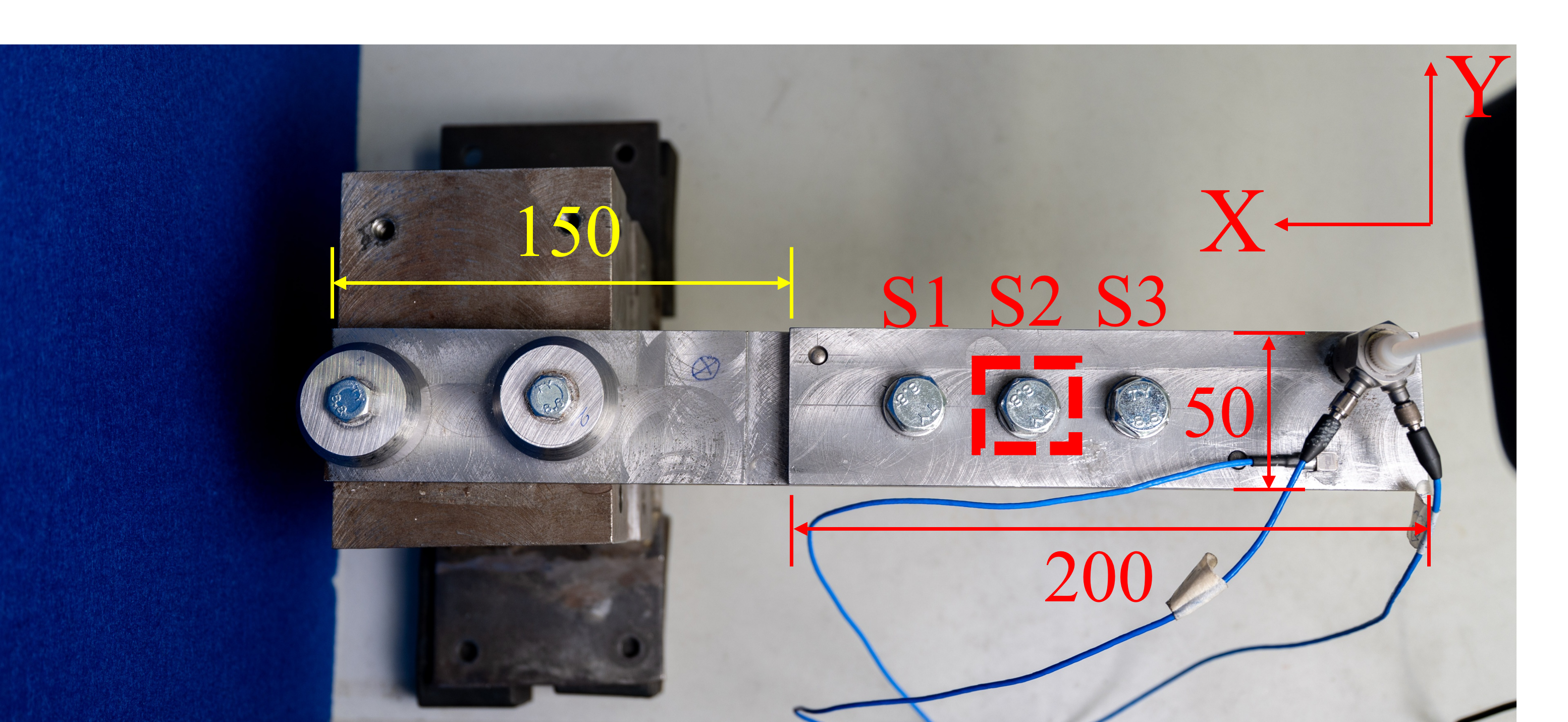}
  \caption{Demonstrator geometry, all dimensions in mm. Beam thickness: 9 mm. Symbols: circle = accelerometer, cross = impact point, square = bolt under test.}
  \label{fig:geometry}
\end{figure}

\section{Results}\label{sec:results}

The bolt under test was tightened to its 80\% preload value to understand the modal behavior of the system in the fully tight position (reference position). Sine-sweep (Fig.~3) and narrow-band white-noise (Fig.~4) inputs were applied to the demonstrator separately using the mechanical shaker.

\begin{figure}[t]
  \centering
  \includegraphics[width=0.98\columnwidth]{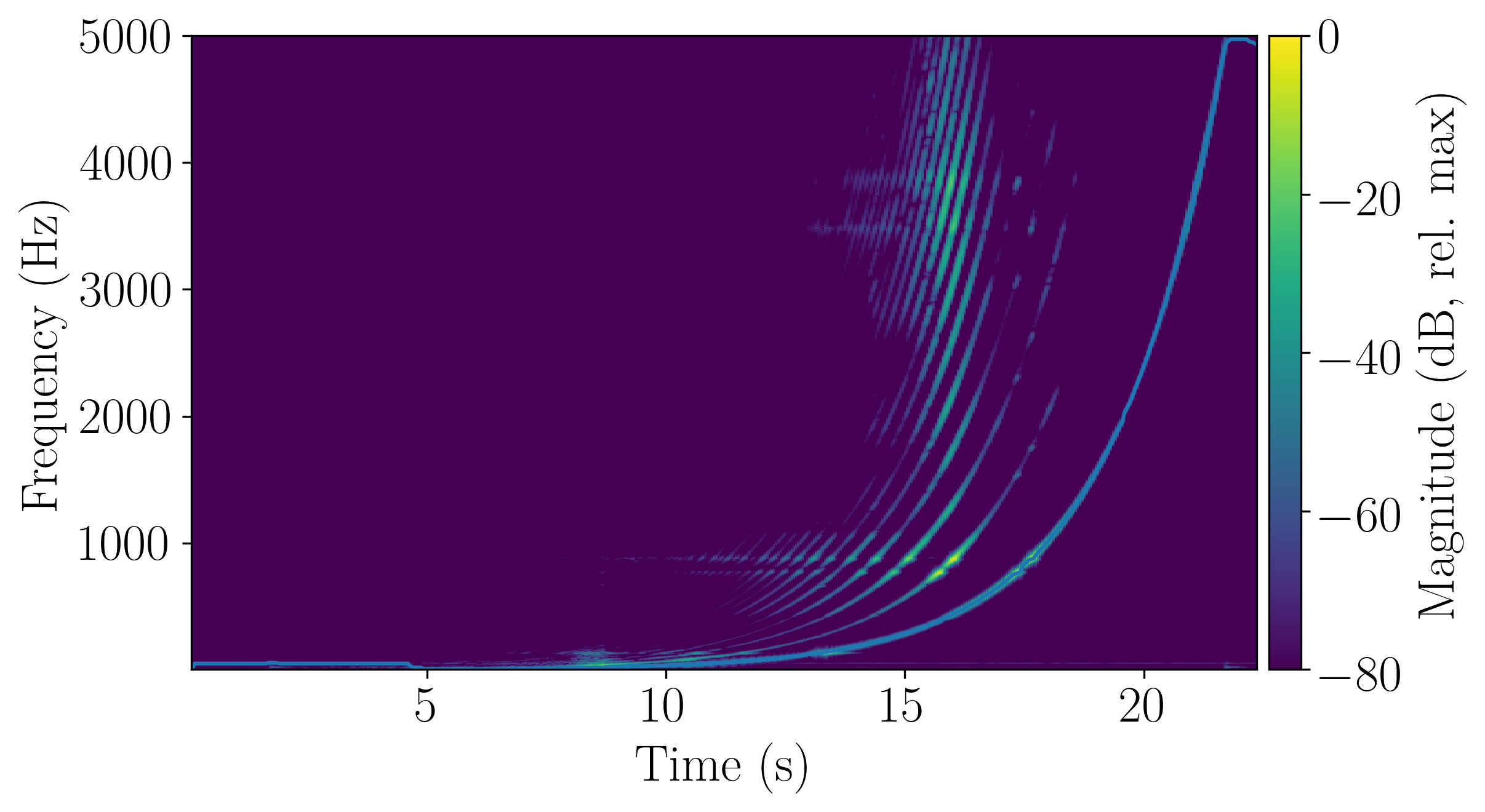}
  \caption{Sine-sweep stimulus applied to the demonstrator, read by PCB 288D01.}
  \label{fig:sinesweep}
\end{figure}

\begin{figure}[t]
  \centering
  \includegraphics[width=0.98\columnwidth]{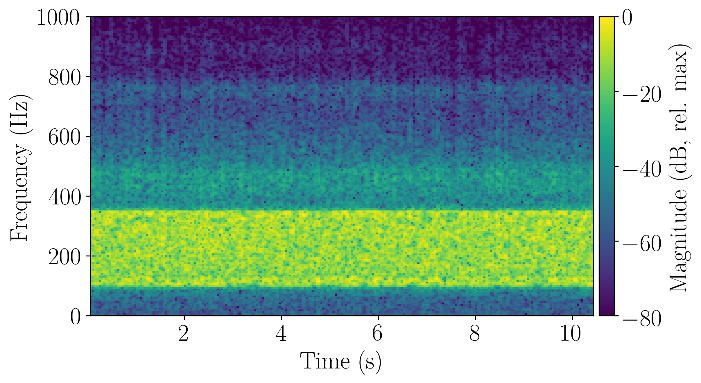}
  \caption{Narrow-band stimulus applied to the demonstrator, read by PCB 288D01.}
  \label{fig:narrowband}
\end{figure}

The sine-sweep signal had frequency components between 1 Hz and 5000 Hz, whereas the narrow-band white noise had a bandwidth of 250 Hz (100 Hz--350 Hz). The results of both trials are given in Fig.~5 and Fig.~6, respectively. Both measurements showed that the demonstrator had a vibrational resonance frequency at $130 \pm 2$ Hz.

\begin{figure}[t]
  \centering
  \includegraphics[width=0.98\columnwidth]{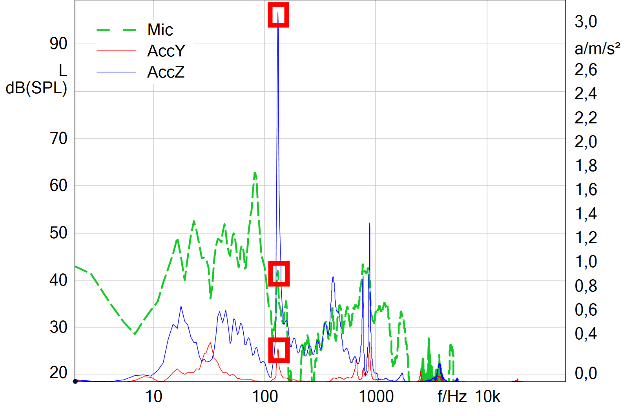}
  \caption{Reaction of the demonstrator under sine-sweep excitation showing clear resonance peaks at several frequency points.}
  \label{fig:sweep_response}
\end{figure}

\begin{figure}[t]
  \centering
  \includegraphics[width=0.98\columnwidth]{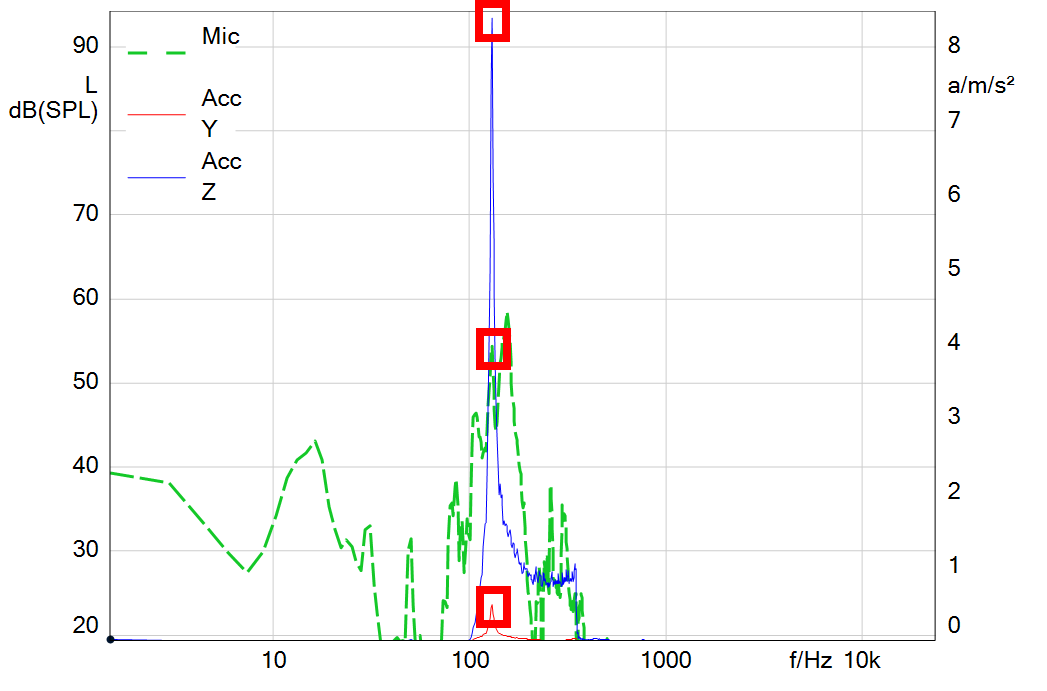}
  \caption{Reaction of the demonstrator under narrow-band white-noise excitation showing clear resonance peaks at several frequency points.}
  \label{fig:noise_response}
\end{figure}

Based on this information, a single 130 Hz sinusoidal vibration input was applied to the demonstrator. The loosened bolt created additional modal peaks in the PSD spectrum. The strongest modal peaks are given in Fig.~7. The additional modal peaks are 1599.5 Hz, 3314.8 Hz, 3919.0 Hz, 5698.9 Hz, and 8451.1 Hz. All peaks were local maxima, had a minimum spacing of 80 Hz from neighbouring modal peaks, and had a prominence of at least 3 dB.

\begin{equation}
  x(t) = A \sin\left( 2\pi f_c t + \beta \sin\left( 2\pi f_m t \right) \right)
  \label{eq:fm}
\end{equation}

FM signals were also applied to the system, taking the 130 Hz resonance frequency point in the middle and iterating over frequency ranges. Frequency-modulated signals were designed using Eq.~\eqref{eq:fm}. The frequencies were modulated between 125 Hz and 135 Hz. The centre frequency was $f_c = 130$ Hz, and the modulation frequency $f_m$ took the values 1 Hz, 2 Hz, 5 Hz, 10 Hz, and 20 Hz.

\begin{figure}[t]
  \centering
  \includegraphics[width=0.98\columnwidth]{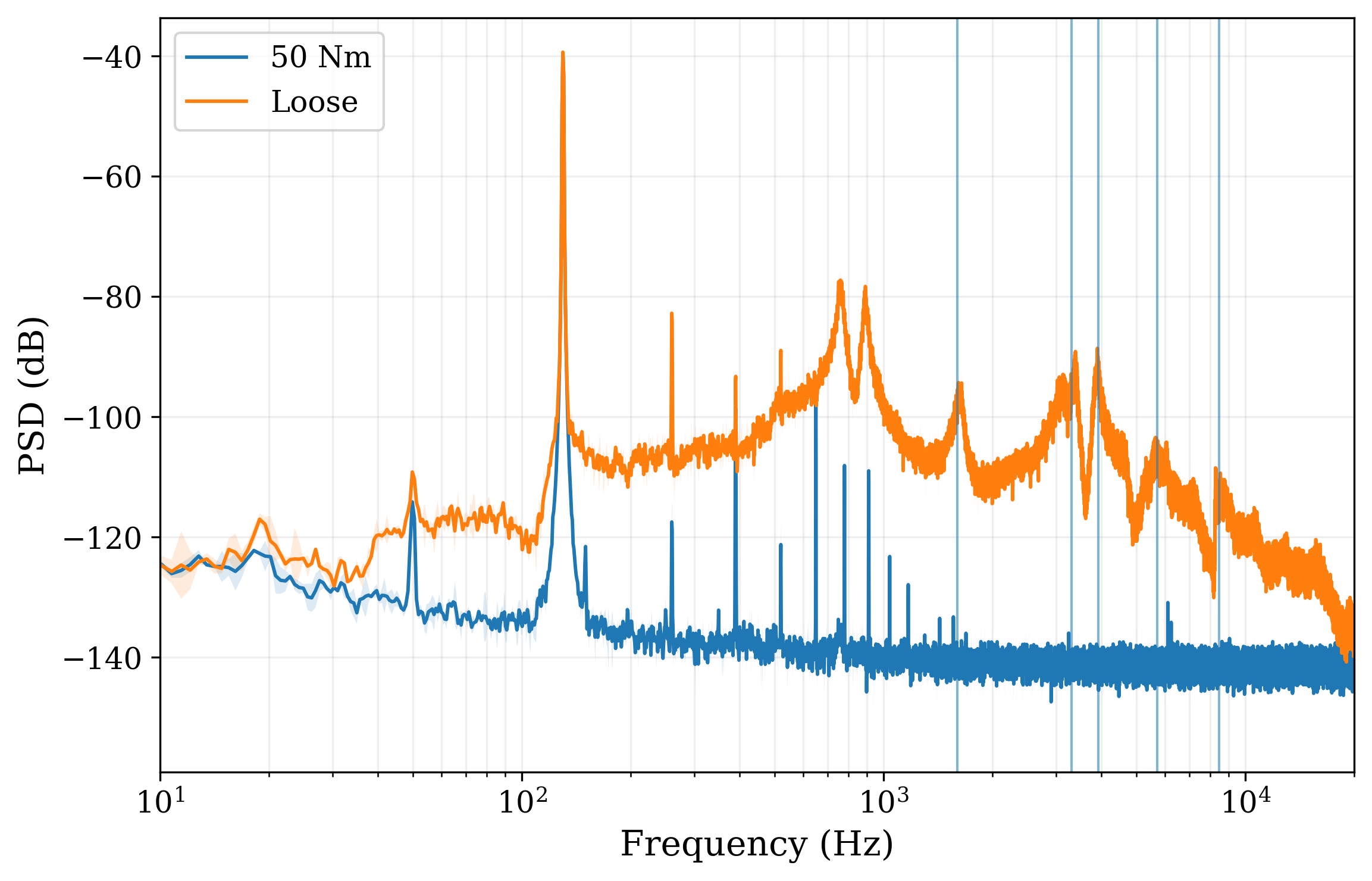}
  \caption{New modal peaks appearing in the loose-bolt measurements are highlighted in light blue. The data is taken from the accelerometer z-axis.}
  \label{fig:loose_peaks}
\end{figure}

From the PSD data of the measurements, meaningful values for separating the loose-bolt state from the tight state were found for $f_m = 2$ Hz. Using Eq.~\eqref{eq:power_ratio},

\begin{equation}
  R_l = 10 \log_{10}\left( \frac{P_{\mathrm{out}}(130\,\mathrm{Hz})}{P_{\mathrm{out}}(l)} \right)
  \label{eq:power_ratio}
\end{equation}

a power ratio between different torque values could be calculated. Here, $P_{\mathrm{out}}(k)$ is the output power around the carrier band near 130 Hz and $P_{\mathrm{out}}(l)$ is the output power around the $l$-th harmonic band. For example, for a frequency-modulated input of 125 Hz--135 Hz, the investigated bands were 250 Hz--270 Hz for the 2nd harmonic and 750 Hz--810 Hz for the 6th harmonic. The results are given in Table~1.

\begin{table}[t]
  \caption{Error bands calculated from three different measurements. All values are given in dB.}
  \label{tab:errorbands}
  \centering
  \begin{tabular}{|l|c|c|}
    \hline
    Preload (percentage) & $l = 2$ & $l = 6$ \\
    \hline
    Loose & $-43.8 \pm 0.4$ & $-21.5 \pm 0.4$ \\
    \hline
    20\%  & $-55.5 \pm 0.1$ & $-53.7 \pm 0.6$ \\
    \hline
    40\%  & $-58.8 \pm 0.0$ & $-53.4 \pm 0.8$ \\
    \hline
    80\%  & $-61.3 \pm 0.1$ & $-58.0 \pm 0.5$ \\
    \hline
  \end{tabular}
\end{table}

\section{Discussion}\label{sec:discussion}

As can be seen from the experiments, the cantilever has a distinct resonance around 130 Hz, which is a good carrier for the forthcoming single-tone and FM excitations. In structures with joints, the preload can cause sticking or slippage of the interface, which can lead either to high effective stiffness with approximately linear response or to lower stiffness with increased frictional loss and strong nonlinearity. This effect is well known to cause shifts in resonance frequencies and changes in damping characteristics because of the dominant effect of the joint on the boundary conditions and energy loss of the assembly. From the modal test results, lower preload is expected to cause shifts in resonance peaks due to changes in stiffness and broadening of the peak due to frictional loss, which are well-known effects of joint conditions in frequency response measurements.

In the context of rail operations, resonance regions play an important role because the carbody and other components are subject to excitation over a large spectrum of frequencies due to irregularities of the rail, wheel--rail interaction, and sudden transient events [5], [8]. Tests such as the self-loosening test according to DIN 65151 or ISO 16130 provide information regarding clamp-force reduction due to controlled transverse motion. However, these tests are not designed to directly access the transfer behaviour of the components, especially the vibro-acoustic transfer, without preload sensors in the actual environment [9]--[11]. As such, the present work is more focused on the practical application of the test results, aiming to identify reliable indicators on the response side, especially in vibration and sound signals, which can be obtained with minimal modification of the components.

One important finding from the single-tone 130 Hz tests is the presence of additional noticeable spectral peak components at higher-frequency bands for the loose-joint case. However, these peak components should be viewed with caution and considered as ``additional or strongly amplified spectral components.'' One possible explanation for their presence is the nonlinear behaviour of the loose interface, which may have acted like a nonlinear element and transferred energy from the driven low-frequency bands to higher bands through the generation of harmonics and broadband components due to frictional micro-slip and intermittent contact between the two parts of the loose interface. The literature on joint dynamics states that frictional contact can introduce higher harmonics and broadband components into the response, which are either absent or very weak during the high-preload, high-linearity state, and that the amplitude of these components increases with the transition from stick to partial-slip and intermittent-contact states [12]--[18]. It is also plausible that, if the loose interface generates impulsive components through repeated loss and re-establishment of contact, these broadband components may couple with higher-frequency structural modes and produce spectral peaks far above the carrier frequency. This rattle-like phenomenon and the increase in noise radiation with decreasing preload are consistent with the broader literature on the nonlinear behaviour of mechanical assemblies and contact nonlinearity.

The concept of harmonic band power ratios provides a clear and easily interpretable means of measuring nonlinear energy transfer. By examining ratios of power in harmonic bands relative to the carrier band, a dimensionless measure is obtained that makes nonlinear joint behaviour easier to interpret. Furthermore, the clear distinction seen in these ratios for increasing harmonic indices, such as in the sixth-harmonic region, reflects the greater difficulty of generating higher harmonics in a nonlinear system. Low-order harmonics are easily generated even with a small amount of nonlinearity, whereas a clear rise in higher harmonics is a more definitive indicator of micro-slip or intermittent impacts in a system [12], [18]. This approach is in line with more general vibro-acoustic modulation and contact acoustic nonlinearity concepts, where a nonlinear contact acts as a modulator and increases in severity with damage [35], [38].

Compared with single-tone vibration, FM excitation offers two advantages. First, sweeping the FM signal roughly through the range 125--135 Hz is more likely to stimulate the general resonance region rather than a single exact point. This improves the chance of capturing resonance shifts and bandwidth changes as preload alters the boundary conditions, which is conceptually similar to general sweep- and frequency-response-function-based modal testing [20], [22]. Second, the time-varying nature of the excitation is more likely to increase the visibility of nonlinear transfer effects and sidebands. In practical shaker application, FM is therefore a simple way to increase sensitivity to nonlinearity without requiring a separate high-frequency probe [35], [38]. However, it is important to note that FM does not itself create looseness signatures; rather, it enhances the visibility of differences between preload states by more fully exciting the resonance region.

\section{Conclusion}\label{sec:conclusion}

This paper proposes a method to assess the looseness of a bolted joint based solely on a shaker device together with an accelerometer and a microphone mounted on a rail-vehicle-inspired setup. It demonstrates that, by providing fixed-frequency and FM excitations around a dominant mode near 130 Hz, loosening effects can be identified through additional peaks in the modal response spectrum and changes in simple power-ratio values between the carrier and higher harmonic frequencies. The method is still at the proof-of-concept stage, with only one bolt considered, several discrete torque levels applied, and only a limited set of features reported. Future work should generalize the method to different geometries, multiple bolts, mounting conditions that better replicate rail-vehicle body and roof structures, and ideally direct preload or clamp-force measurements.

\section{Acknowledgements}\label{sec:acknowledgements}

This work was funded by the European Regional Development Fund (ERDF/EFRE) and the State of Brandenburg within the StaF-Verbund programme, under the project ``Systementwicklung fur intelligente und automatisierte mobile Inspektion von Schienenfahrzeugen (SIAMIS)'', administered by the Investment Bank of the State of Brandenburg (ILB).

\section*{References}
\small
[1] National Aeronautics and Space Administration: ``Fastener Design Manual,'' NASA Reference Publication 1228, 1990.\par
[2] H. Gong: ``Review of research on loosening of threaded fasteners,'' \emph{Friction}, vol.~10, no.~3, pp.~335--359, 2022. DOI: 10.1007/s40544-021-0497-1.\par
[3] Deutsches Institut fur Normung: DIN 25201-4:2021-11, ``Design guideline for railway vehicles and their components--Bolted joints--Part 4: Securing of bolted joints,'' 2021.\par
[4] G.~H. Junker: ``New criteria for self-loosening of fasteners under vibration,'' SAE Technical Paper 690055, 1969.\par
[5] Deutsches Institut fur Normung: DIN EN 12663-1:2024-02, ``Railway applications - Structural requirements of railway vehicle bodies - Part 1: Locomotives and passenger rolling stock (and alternative method for freight wagons),'' 2024.\par
[6] A. Orvnas: ``Methods for Reducing Vertical Carbody Vibrations of a Rail Vehicle - Literature Survey,'' KTH Royal Institute of Technology, report, 2010.\par
[7] A. Mosleh \emph{et al.}: ``Recent advances in wayside condition monitoring for railways: a comprehensive review,'' \emph{Railway Engineering Science}, 2026. DOI: 10.1007/s40534-025-00423-2.\par
[8] I. Bravo, U. Alvarado, J. Nieto, and P. Ciaurriz: ``On-board accelerometers in railway track condition monitoring. A systematic review,'' \emph{Transportation Research Interdisciplinary Perspectives}, vol.~33, 101572, 2025. DOI: 10.1016/j.trip.2025.101572.\par
[9] Deutsches Institut fur Normung: DIN 65151:2002-08, ``Dynamic testing of the locking characteristics of fasteners under transverse loading conditions (vibration test),'' 2002.\par
[10] B. Pichoff: ``Bolt self-loosening, an old problem revisited,'' \emph{Materiaux \& Techniques}, vol.~106, art.~607, 2018. DOI: 10.1051/mattech/2018034.\par
[11] International Organization for Standardization: ISO 16130:2015, ``Aerospace series--Dynamic testing of the locking behaviour of bolted connections under transverse loading conditions (vibration test),'' 2015.\par
[12] Y. Wang \emph{et al.}: ``Experimental studies on the energy dissipation of bolted structures with frictional interfaces: A review,'' \emph{Friction}, 2024. DOI: 10.1007/s40544-023-0809-8.\par
[13] D.~O. Smallwood: ``Damping investigations of a simplified frictional shear joint,'' technical report, 2000.\par
[14] M. Ahmadian and H. Jalali: ``Identification of bolted lap joints parameters in assembled structures,'' \emph{Mechanical Systems and Signal Processing}, vol.~21, no.~2, pp.~1041--1050, 2007.\par
[15] N. Jamia, H. Jalali, J. Taghipour, M.~I. Friswell, and H.~H. Haddad Khodaparast: ``An equivalent model of a nonlinear bolted flange joint,'' \emph{Mechanical Systems and Signal Processing}, vol.~153, art.~107507, 2021. DOI: 10.1016/j.ymssp.2020.107507.\par
[16] M. Krack \emph{et al.}: ``The Tribomechadynamics Research Challenge: Confronting blind predictions for the linear and nonlinear dynamics of a thin-walled jointed structure with measurement results,'' \emph{Mechanical Systems and Signal Processing}, vol.~224, art.~112016, 2025. DOI: 10.1016/j.ymssp.2024.112016.\par
[17] A.~A. Morsy and P. Tiso: ``Predicting the variability of the dynamics of bolted joints using polynomial chaos expansion,'' \emph{Mechanical Systems and Signal Processing}, 2025.\par
[18] D. Bograd, P. Reuss, A. Schmidt, L. Gaul, and M. Mayer: ``Modeling the dynamics of mechanical joints,'' \emph{Mechanical Systems and Signal Processing}, pp.~2801--2826, 2011.\par
[19] N. Li \emph{et al.}: ``Monitoring of bolt looseness using piezoelectric transducers: Three-dimensional numerical modeling with experimental verification,'' \emph{Journal of Intelligent Material Systems and Structures}, 2020. DOI: 10.1177/1045389X20906003.\par
[20] D.~J. Ewins: \emph{Modal Testing: Theory, Practice and Application}, 2nd ed. Baldock, UK: Research Studies Press, 2000.\par
[21] Brüel \& Kjaer: ``Mechanical Mobility Measurements,'' Technical Review BR 0458-012, 2012.\par
[22] Brüel \& Kjaer: ``Frequency Response Function Measurements: Shaker Excitation,'' Application Note BO 0419-11, 2012.\par
[23] T. Eraliev \emph{et al.}: ``Vibration-Based Loosening Detection of a Multi-Bolt Structure Using Machine Learning Algorithms,'' \emph{Sensors}, vol.~22, no.~3, art.~1210, 2022. DOI: 10.3390/s22031210.\par
[24] N. Nicholas, J. Wilson, and C.~J. Stowell: ``Feasibility of using low-sampled accelerometer measurements for bolt joint looseness detection,'' \emph{IET Renewable Power Generation}, vol.~16, no.~12, pp.~2762--2777, 2022.\par
[25] I. Ho, A.~M. Boor, S. Vanlanduit, and J.~V. Dirckx: ``Bolt looseness detection using Spectral Kurtosis analysis for structural health monitoring,'' in \emph{Proc. ISMA/USD}, 2020.\par
[26] S.~M. Sah, J.~J. Thomsen, M. Brons, and N.~A. Jorgensen: ``Estimating bolt tightness using transverse natural frequencies,'' \emph{Journal of Sound and Vibration}, vol.~431, pp.~137--149, 2018. DOI: 10.1016/j.jsv.2018.05.040.\par
[27] L.-Y. Hou and D.-B. Zhuo: ``Research on bolt loosening recognition based on sound signal and GA-SVM-RFE,'' \emph{Scientific Reports}, vol.~15, art.~24165, 2025. DOI: 10.1038/s41598-025-08730-8.\par
[28] L. Cheng \emph{et al.}: ``Sound Sensing: Generative and Discriminant Model-Based Approaches to Bolt Loosening Detection,'' \emph{Sensors}, vol.~24, no.~19, art.~6447, 2024. DOI: 10.3390/s24196447.\par
[29] Y. Zhou \emph{et al.}: ``Percussion-based bolt looseness identification using vibration-guided sound reconstruction,'' \emph{Structural Control and Health Monitoring}, vol.~29, no.~2, e2876, 2022. DOI: 10.1002/stc.2876.\par
[30] Y. Zhang \emph{et al.}: ``Bolt loosening detection based on audio classification,'' \emph{Advances in Structural Engineering}, vol.~22, pp.~2882--2891, 2019. DOI: 10.1177/1369433219852565.\par
[31] F. Wang \emph{et al.}: ``Bolt-looseness detection by a new percussion-based method using multifractal analysis and gradient boosting decision tree,'' \emph{Structural Health Monitoring}, 2020. DOI: 10.1177/1475921720912780.\par
[32] Z. Yang \emph{et al.}: ``Bolt preload monitoring based on percussion sound signal and convolutional neural network (CNN),'' \emph{Nondestructive Testing and Evaluation}, 2022. DOI: 10.1080/10589759.2022.2030735.\par
[33] Z. Jiang \emph{et al.}: ``Identification of Bolt Loosening Damage of Steel Truss Structure Based on MFCC-WPES and Optimized Random Forest,'' \emph{Applied Sciences}, vol.~14, no.~15, art.~6626, 2024. DOI: 10.3390/app14156626.\par
[34] R. Yuan, Y. Lv, Q. Kong, and G. Song: ``Percussion-based bolt looseness monitoring using intrinsic multiscale entropy analysis and BP neural network,'' \emph{Smart Materials and Structures}, vol.~28, art.~125001, 2019. DOI: 10.1088/1361-665X/ab3b39.\par
[35] Z. Zhang, M. Liu, Y. Liao, Z. Su, and Y. Xiao: ``Contact acoustic nonlinearity (CAN)-based continuous monitoring of bolt loosening: Hybrid use of high-order harmonics and spectral sidebands,'' \emph{Mechanical Systems and Signal Processing}, vol.~103, pp.~280--294, 2018. DOI: 10.1016/j.ymssp.2017.10.009.\par
[36] H. Gong, J. Huang, J. Liu, and X. Deng: ``Proof-of-concept study of high-order sideband for bolt loosening detection using vibroacoustic modulation method,'' \emph{Mechanical Systems and Signal Processing}, vol.~169, art.~108638, 2022. DOI: 10.1016/j.ymssp.2021.108638.\par
[37] N. Zhao, L. Huo, and G. Song: ``Vibration acoustic modulation for bolt looseness monitoring based on frequency-swept excitation and bispectrum,'' \emph{Smart Materials and Structures}, vol.~32, art.~034004, 2023. DOI: 10.1088/1361-665X/acb579.\par
[38] J.~C. Pineda Allen and C.~T. Ng: ``Nonlinear Guided-Wave Mixing for Condition Monitoring of Bolted Joints,'' \emph{Sensors}, vol.~21, no.~15, art.~5093, 2021. DOI: 10.3390/s21155093.

\normalsize

\end{document}